\documentclass[twocolumn,printnumbers,amsmath,amssymb,showpacs,10pt]{revtex4}

\newcommand{\be}{\begin{equation}}
\newcommand{\ee}{\end{equation}}
\newcommand{\ba}{\begin{eqnarray}}
\newcommand{\ea}{\end{eqnarray}}
\usepackage{graphicx}
\usepackage{dcolumn}
\usepackage{bm}
\usepackage{setspace}
\usepackage{color}
\begin{document}

\title{Geometric background charge: dislocations  on capillary bridges.}
\author{W. T. M. Irvine $^{*}$ and V. Vitelli$^{\dag}$}
\affiliation{$^{*}$ James Franck Institute, 929 E 57th Street, Chicago, IL 60637, USA \\
$^{\dag}$ Instituut-Lorentz for Theoretical Physics, Universiteit Leiden, 2300 RA Leiden, The Netherlands
}

\begin{abstract} 
\noindent 
Recent experiments have shown that colloidal crystals confined to weakly curved capillary bridges introduce groups of dislocations organized into  `pleats' as means to relieve the stress caused by the Gaussian curvature of the surface. We consider the onset of this curvature-screening mechanism, by examining the energetics of isolated dislocations and interstitials on capillary bridges with free boundaries. The  boundary provides an essential contribution to the problem, akin to a background charge that ``neutralizes'' the unbalanced integrated  curvature of the surface. This makes it favorable for topologically neutral dislocations and groups of dislocations - rather than topologically charged disclinations and scars - to relieve the stress caused by the unbalanced gaussian curvature of the surface. This effect applies to any crystal on a surface with non-vanishing integrated Gaussian curvature and stress-free boundary conditions. 
 We corroborate the analytic results by numerically computing the energetics of a defected lattice of springs confined to surfaces with weak positive and negative curvature.\end{abstract}

\maketitle

When a hexagonal lattice is draped over a curved substrate, its bonds necessarily stretch or compress to accommodate the underlying curvature \cite{NelsonBook,paulbook}. This phenomenon, known as geometric frustration, arises from the competition between local particle interactions, that favor an ordered lattice structure, and geometric constraints that prevent local order from filling space \cite{nelsonpeliti,BNT:2000,vv,RMPV,RMPR,Santa,BG:2009,mark,mark2,Hex}. An intriguing connection has recently emerged between the frustrated order of twisted filament assemblies and the apparently distinct problem of curved crystalline order \cite{Greg}.

\begin{figure}[!ht]
\centering
\includegraphics[width=0.8 \columnwidth]{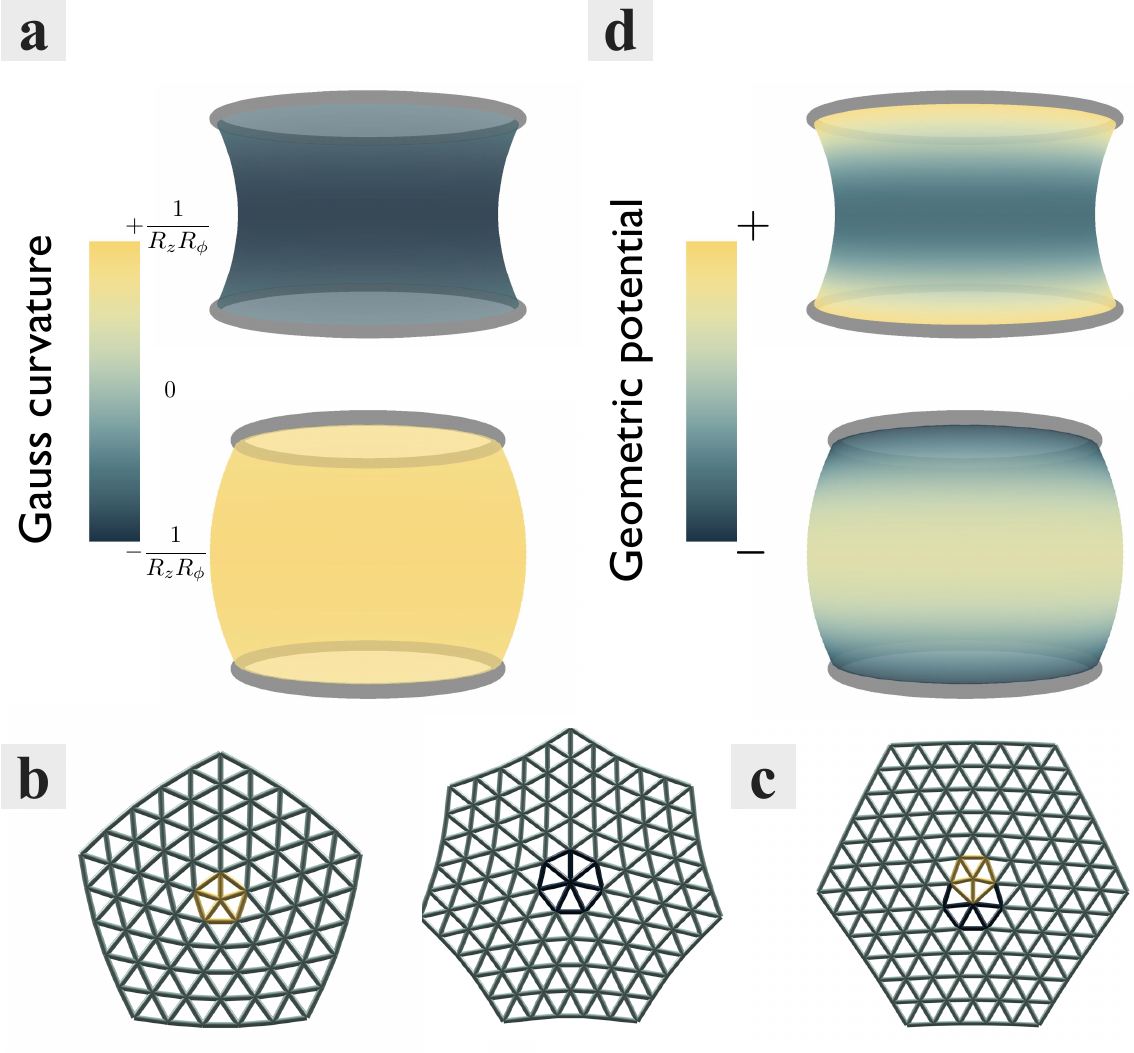}
\caption{ {\bf (a)} Profiles of capillary bridges coloured by the magnitude of the Gaussian curvature on their surface (Eq.~\ref{eq:metric3}). For the weakly curved capillary bridges shown here,  the curvature   $G\sim \frac{1}{R_z R_\phi}$  is approximately uniformly negative or positive. $R_z$ and $R_\phi$ are the radii of curvature in the vertical and azimuthal direction respectively. 
{\bf  (b) } Disclinations  in a hexagonal lattice appear as five or seven-fold coordinated particles and  carry an angular topological charge that can compensate the angular deficit created by  the Gaussian curvature. {\bf (c)} Dislocations in a hexagonal lattice appear as dipoles of disclinations. The topological charge of disclinations appears in Eq. \ref{eq:vonKarman} as a source of geometric stress on the same footing as Gaussian curvature,  suggesting a direct defect-curvature coupling, with five(seven)-fold disclinations coupled to positive(negative) curvature.   The energetics of disclinations are  however  more closely analogous to those of an electric charge that interacts with a geometric `charge density' $\rho$, shown in panel (d). {\bf (d)} $\rho$  is the  solution to a Laplacian equation (Eq. \ref{eq:metricw}) with $G$ as its source and   an additional   constant term $\rho_{{\rm H}}$ that originates from the stress-free boundary condition. In the case of surfaces with unscreened curvature and free boundaries, this term is akin to a  background charge that ÒneutralizesÓ the unbalanced integrated curvature of the surface and leads to an energetic  preference for the appearance of dislocations (dipoles) over disclinations, despite the uniform Gaussian curvature.}
\label{spherical_xtal}
\end{figure}

A striking experimental realization of curved space crystallography is provided by colloidal monolayers confined on spherical droplets \cite{BB2003,NatMat2005} and on capillary bridges of varying positive or negative curvature \cite{IVC:2010}, see Fig. 1a. In both cases, the curvature-generated mechanical strains are screened by defects in the crystal lattice, especially dislocations  or disclinations (see Fig. 1b,c).

In a spherical crystal, there must be an excess of twelve five-fold or (positive) disclinations simply because you cannot tile a sphere with hexagons, as demonstrated by a classic soccer ball. The presence of these twelve excess disclinations is dictated by topology, energetics only fixes their positions at the vertexes of an icosahedron inscribed in the sphere. The experimental and theoretical studies of Ref. \cite{BB2003,BNT:2000} have demonstrated that introducing additional grain boundary "scars" emanating from each of the twelve positive disclinations becomes energetically favourable on large spheres, even if they are not required by topology.

Defect nucleation on curved capillary bridges shaped as catenoids or barrels works differently than on the sphere: there are no topological requirements that force isolated disclinations in the ground state. Instead defects in the form of isolated dislocations are first nucleated to screen the curvature. Do "scars" exist on these surfaces even without the isolated disclinations from which they emanate in spherical crystals? Recent experiments show clear evidence for the existence of finite length low-angle grain boundaries that do not originate from free disclinations \cite{IVC:2010}. These defect structures play an analogous role to pleats in fabrics - they accommodate for curvature  by inserting and terminating extra rows of particles.

\begin{figure*}[!t]
\centering
\includegraphics[width=2 \columnwidth]{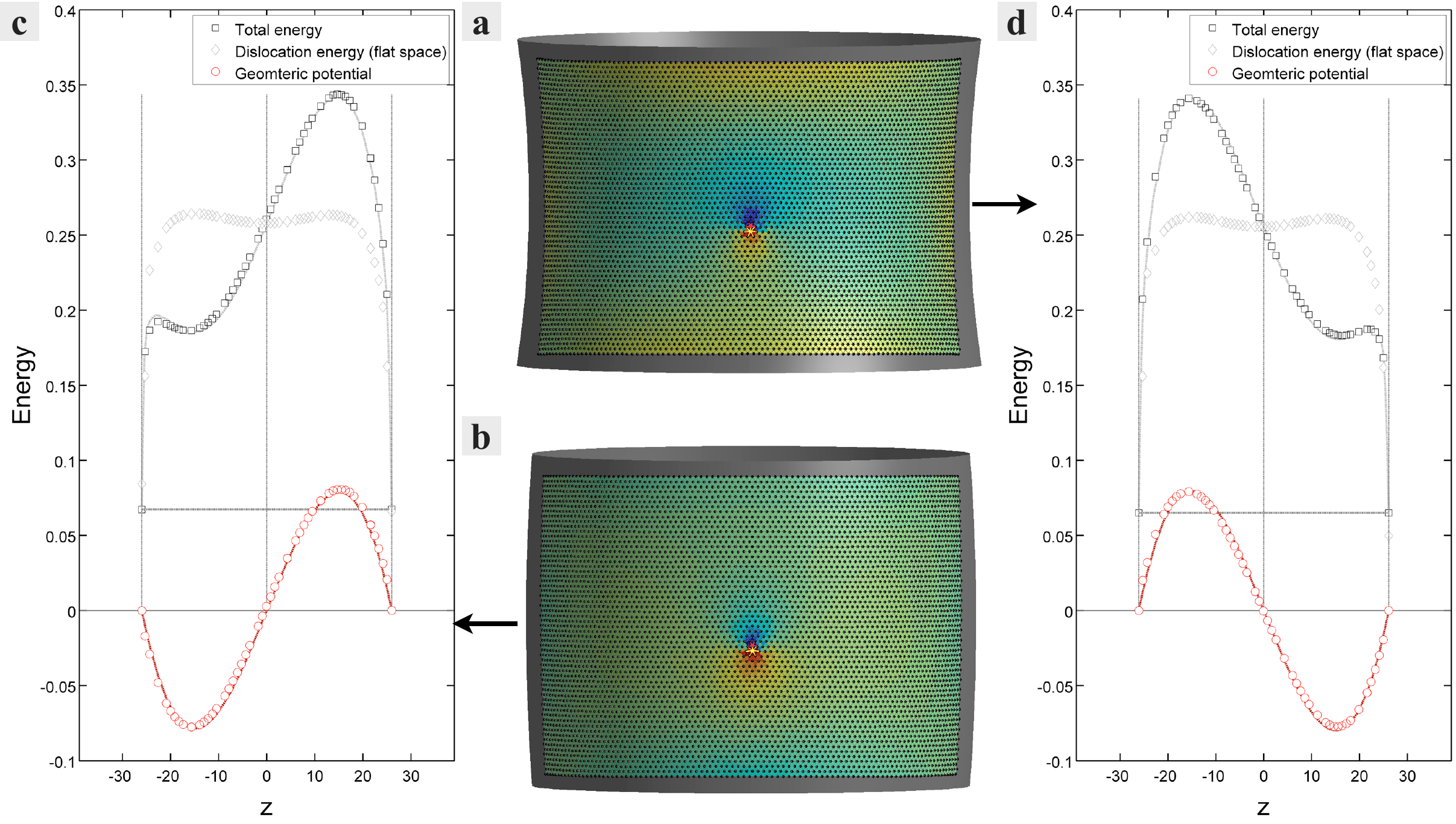}
\caption{Energy (c,d) of a defected lattice of springs of unit rest length and unit spring constant, constrained to lie on the surface of capillary bridges (a,b). 
A dislocation, polarized so its five-fold particle lies above its seven-fold particle was inserted in the crystalline patch at positions of varying height in order to map out the potential felt by an isolated dislocation.
The minimum energy of each configuration was found using a conjugate-gradient minimization similar to that used in ref~\cite{vv} and is plotted (hollow squares in  c(d)) for the capillary bridges shown in b(a).  The flat-space energies (grey diamonds  in c(d)) and energy of geometric frustration (black solid line  in c(d)) of an un-defected patch are subtracted in (c,d)  to reveal (red hollow circles) the predicted geometric potential (solid red line). }
\label{spherical_xtal}
\end{figure*}

The building block of `pleats' is dislocations. In this article we present an analytical and numerical study of the energetics of dislocations on capillary bridges of positive and negative Gaussian curvature. Our results highlight
the ability of dislocations to screen curvature and provide an accurate estimate of the forces acting on them as a result of the simultaneous presence of curvature and free boundaries. As we shall see, of crucial importance is a subtle geometric force exerted on an isolated dislocation by a neutralizing "background charge" that  counter-balances the effect of a non-vanishing integrated Gaussian curvature of the surface, thereby enforcing  the stress-free boundary condition \cite{vv}.

The elastic energy of a curved crystalline monolayer draped on a curved substrate can be written as \cite{NelsonBook,paulbook} 

\begin{equation}
 F=\frac{1}{2Y}  \int  {\rm d}A \ \left( \nabla^2 \chi \right)^2  ,
\label{eq:chi}
\end{equation}
where $Y$ is the Young modulus and $\chi(x)$ is the Airy stress function that satisfies the biharmonic equation sourced by the Gaussian curvature and the disclinations
\begin{equation}
\frac{1}{Y} \nabla^{4} \chi (x)= \sum_\alpha q_\alpha \delta (x-x_\alpha) -G(x) \quad \!\!\! ,
\label{eq:vonKarman}
\end{equation}
where $q_{\alpha}$ and $x_\alpha$ are the disclination charges and positions respectively.

The energetics of these elastic systems can be mapped onto a simpler electrostatic problem. Denote by $\chi^G(x)$ the solution of Eq. (\ref{eq:vonKarman}) in the absence of defects. The biharmonic equation can be solved in two steps \cite{vv,BG:2009}. First introduce an auxiliary function $\rho(x)$ that satisfies:
\begin{equation}
\label{eq:metricw}
\nabla^2 \rho(x) = G(x) \!\!\!\! \quad .
\end{equation}
The second step is to set $\rho(x)$ as a source in the Poisson equation for $\chi^{G}(x)$
\begin{equation}
\label{eq:metricw1}
\frac{1}{Y} \nabla^2 \chi^G (x)= - ( \rho(x) - \rho_H(x)) \!\!\!\! \quad  ,
\end{equation}
where $ \rho_H(x) $ is  a harmonic function that we will use to satisfy the boundary condition of vanishing stress at the edge of the capillary bridge \cite{vv,BG:2009}. Note that the geometric charge density $\rho(x)$ is analogous to a smeared out electrostatic charge that, as we shall see, interacts with the defects.

In order to enforce the boundary condition, we need to determine the stress tensor $\sigma_{ij}(x)$ in an arbitrary set of curvilinear coordinates $x=\{x_1,x_2\}$ which can be expressed in terms of $\chi(z)$ and the metric tensor $g_{ij}$ as \cite{Tensor_book}:
\begin{equation}
\label{eq:stress1}
\sigma_{11} = \frac{\tau_{11}}{g_{11}} \qquad 
\sigma_{22} = \frac{\tau_{22}}{g_{22}} \qquad
\sigma_{12} = \frac{\tau_{12}}{\sqrt{g_{11} g_{22}}} ,
\end{equation}
where
\begin{equation}
\label{eq:stress2}
\tau_{ij} \equiv \gamma^{m n} \gamma_{j p} g_{im} g^{p r} \left( \partial_n \partial_r \chi(z) - \Gamma^q_{nr} \partial_q \chi(z) \right) \!\!\!\! \quad .
\end{equation}
The Christoffel symbols are denoted by $\Gamma^k_{ij}$ and $\gamma_{ij}=\epsilon_{ij}/\sqrt{g}$ is the covariant antisymmetric tensor, with the determinant of the metric $g=g_{11} g_{22}$ . 

To facilitate comparison with our experimental studies, we focus on capillary bridges that are surfaces of revolution whose radius $r$ is approximately given as a function of the height $z$ by:
\begin{equation}\label{eq:metric6}
r(z) =  R_\phi + R_z - R_z \cosh{\left[\frac{z}{R_z}\right]} \!\!\!\! \quad
\end{equation}
This choice describes both the catenoids and barrels of Fig. 1a. $R_\phi$ is always positive, while  $R_z$ is positive on barrels and negative on catenoids.  In both cases, the resulting  metric reads:
\begin{equation}\label{eq:metric}
ds^2= \cosh{\left[\frac{z}{R_z}\right]}^2 \!\!\!\!\!\! \quad dz^2 + \left(R_\phi + R_z - R_z \cosh{\left[\frac{z}{R_z}\right]} \!\!\!\! \quad \right)^2 \!\!\!\!\!\! \quad d\phi^2
\end{equation}
We assume that the total height of our catenoids and barrels is equal to $H$, so that $-\frac{H}{2}<z<+\frac{H}{2}$.
The Gaussian curvature of these surfaces is given by:
\begin{equation}\label{eq:metric3}
G(z) = \frac{{\rm sech}^3\left(\frac{z}{R_z}\right)}{R_z (R_\phi-R_z)+R_z^2 \cosh \left(\frac{z}{R_z}\right)}
\end{equation}
For the capillary bridges, we can calculate $\sigma_{zz}(z)$ and  $\sigma_{\phi\phi}(z)$ from Eqs. (\ref{eq:stress1}-\ref{eq:stress2}) giving:
\begin{equation}\label{eq:GB2}
\sigma_{zz}(z)=\frac{{\rm sech} \left(\frac{z}{R_z}\right) \tanh \left(\frac{z}{R_z}\right)}{ (R_\phi-R_z)+ R_z \cosh \left(\frac{z}{R_z}\right)} \ \ \partial_z \chi \!\!\!\! \quad ,
\end{equation}
and
\begin{equation}\label{eq:GB2b}
\sigma_{\phi\phi}(z)=\frac{1}{\cosh^2{(\frac{z}{R_z})}}\left[\partial_z^2 \chi -\frac{1}{R_z} \tanh{(\frac{z}{R_z})} \!\!\!\! \quad \partial_z \chi \right]  .
\end{equation} 
The sign convention we adopt for $\sigma_{ij}$ is that the stress is negative if the material element is compressed and positive if it is stretched. Intuitively we expect that when $z \approx 0$, near the neck of the catenoid (barrel), the applied stress is negative (positive).  

As we shall see, defect nucleation first occurs in the regime of small curvature $H/R_z\ll 1$, for which the capillary bridge can be viewed as a weakly deformed cylinder of radius approximately equal to $R_\phi$ and height $H$. In this regime, we can Taylor expand all previous expressions in powers of $z/R_z$ to find a closed-form solution for $\chi^{G}(z)$ in Eq. (\ref{eq:metricw1}), subject to the free boundary condition \cite{vv} $\sigma^{G}_{zz}(z=\pm H/2)=0$. 
The result reads:
\begin{equation}\label{eq:a8}
 \chi^{G}(z) =  - \frac{Y }{24 R_z R_{\phi}}\left( z^4-\frac{H^2 z^2}{2}+ \frac{H^4}{16} \right)  \!\!\!\! \quad ,
\end{equation}
where the last (constant) term, which is of no physical consequence, was chosen to set $\chi^{G}=0$ on the boundary. Upon plugging Eq. (\ref{eq:a8}) into Equations (\ref{eq:GB2}) and (\ref{eq:GB2b}), we find (to leading order in $H/R_z$) that the stress $\sigma^G_{zz} \approx 0$ while $\sigma^G_{\phi \phi}\approx \frac{Y H^2}{24 R_z R_{\phi}}$ near the neck $z\approx 0$. These results mean that there is spring compression for catenoids ($R_z < 0$ and $\sigma^G_{\phi \phi}<0$) and stretching for barrels ($R_z >0$ and $\sigma^G_{\phi \phi}>0$). Note that the azimuthal component of the stress of geometric frustration, $\sigma^G_{\phi \phi}(z)$, changes sign for $z>H/{\sqrt{12}}$, a fact of crucial importance to determine the geometric force exerted on dislocations by the curvature of the capillary bridge.

According to standard elasticity theory, a dislocation in an external stress field, $\sigma_{ij}(x)$, experiences a Peach-Koehler force, $\vec{f}(x)$, given by \cite{HL:1968} 
\begin{eqnarray}
f_{k}(x)=\epsilon_{kj}b_{i} \sigma_{ij}(x) ,
\label{eq:ac5}
\end{eqnarray}
where $\vec{b}$ is the Burgers vector of the dislocation. 
Similarly, a dislocation introduced into the curved 2D crystal will experience a Peach-Koehler force as a result of the pre-existing stress field of geometric frustration $\sigma^{G}_{ij}(x)$.
If we choose $\vec{b}$ along its minimum-energy orientation for a catenoid (azimuthal, corresponding to a dislocation having its 7-fold defect closer to the neck), we obtain a Peach-Koehler force $f_z(z)$ 
that points in the z direction and has magnitude given by:
\begin{eqnarray}
\label{eq:a5}
f_z(z)= - b \sigma^{G}_{\phi\phi}(z)  \!\!\!\! \quad .
\end{eqnarray}
An explicit formula for $f_z(z)$ on capillary bridges is obtained by plugging $\sigma_{\phi \phi}\approx \partial_z^2 \chi^{G}(z)$ into Eq. (\ref{eq:a5}) with the result
\begin{equation}\label{eq:aw1}
f_z(z)   \approx  - \frac{Y b}{2 R_z R_{\phi}}\left( \frac{H^2 }{12} - z^2\right)  \!\!\!\! \quad .
\end{equation}
For a dislocation with its {\it negative} disclination facing the neck of a capillary bridge with negative curvature, the force is repulsive (attractive), if it is located at a distance $z$ smaller (greater) than $H/\sqrt{12}$. For a barrel, the prefactor $R_z$ switches sign, hence the minimum energy orientation for the dislocation is obtained for $b<0$, eg. the {\it positive} disclination is facing the neck. 

The corresponding potential is obtained upon integrating along the z direction (ignoring higher order metric effects) with the result
\begin{equation}\label{eq:aw2}
\phi(z)   \approx   - \frac{Y b}{6 R_z R_{\phi}}\left(z^3 - \frac{H^2 }{4} z\right)  \!\!\!\! \quad .
\end{equation}
Around the neck $z\ge0$, the second term in Eq. (\ref{eq:aw2}) dominates and the potential $\phi(z)\approx \frac{Y b H^2}{24 R_z R_{\phi}}z$ is repulsive for the catenoid (Fig. 1d) and attractive for the barrel (Fig. 1c). For $z<0$
the dislocation acquires the opposite burger vector and the potential flips sign as expected given that the disclination closer to the neck has now the opposite sign.    

The second term in Eq. (\ref{eq:aw2}) is obtained from fixing the boundary conditions and can be traced all the way to the additional source $\rho_H$ in the Poisson equation (Eq.~\ref{eq:metricw1}). 
It is a contribution to the geometric force experienced by the dislocation that is linear in the Burgers vector and it is induced by the simultaneous presence of both the Gaussian curvature and the boundary. It should not be confused with the interaction of the dislocation with its own image which would be present also on a finite cylindrical patch. In the electrostatic analogy, the source $\rho_H$ subtracted from $\rho(x)$ can be viewed as a neutralizing "background charge" which ensures that the area integral of the right hand side of Eq. (\ref{eq:metricw1}) vanishes. This condition forces the radial stress to vanish at the boundary.

In order to test our analysis, we have performed constrained energy minimizations of a hexagonal lattice of harmonic springs with the dislocation fixed at position $z$ from the center of a patch of a catenoid (Fig. 2a) and a barrel (Fig. 2b). The gray hollow diamonds represent the energy of a patch containing a dislocation at different $z$ on a cylinder of the same size as the catenoid and barrel. It is very flat until the dislocation is placed close to the boundary and starts being attracted by its own image. This flat space contribution to the energetics purely depends on the presence of a boundary. We now subtract from the total energy (black hollow squares) the energy of the same patch in flat space (gray hollow diamonds) and the energy of geometric frustration of an un-defected patch (black solid line) and obtain (red hollow circles)  for both catenoids and barrels the geometric potential, in agreement  (red solid line) with that calculated in Eq. (\ref{eq:aw2}). The agreement with our analytical result (continuos line) that captures all curvature contributions to the energetics is very good.   

Figure 3b shows a systematic comparison of our numerical results (symbols) with the theoretical predictions (continuos lines) based on Eq. (\ref{eq:aw2}) for the $z>0$ portion of the geometric potential of a family of catenoids with increasing curvature. Figure 3a compares a plot of the minimum of $\phi(z)$ vs. $G=\frac{1}{R_zR_{\phi}}$ (the Gaussian curvature at the neck) determined numerically with the theoretical prediction. The agreement is very good, corroborating our analysis for a wider range of curvature than expected on the basis of our perturbative treatment that describes catenoids and barrels as small deformations from a cylindrical geometry. The intuitive reason why the minimum of $\phi(z)$ is located at $z \approx H/{\sqrt{12}}$, where $\sigma_{\phi\phi}$ changes sign, can be explained as follows. If the negative (positive) disclination is facing the neck of the catenoid (barrel), the extra row of atoms emanating from the dislocation is directed towards the $z<z_{min}$ ($z>z_{min}$) portion of the crystal that is stretched -- this choice insures that the elastic energy is minimized. 



We can now use an electrostatic analogy to first determine the geometric potential of a tight dipole composed of a positive and a negative disclination and then verify that it indeed matches our geometric potential for a dislocation obtained analytically in Eq. (\ref{eq:aw2}) and numerically in Fig. 3. The energy of $N$ disclinations labelled by the index $\alpha$ and with ``topological charges'' $\{q_{\alpha}=\pm \frac{2 \pi}{6}\}$ reads \cite{BG:2009,BNT:2000}
\begin{eqnarray}
F =   \frac{Y}{2} \int \! dA \int \! dA'   \!\!\!\! \quad {\cal N}(x)  \!\!\!\! \quad  \frac{1}{\Delta_{xx'}^2} \!\!\!\! \quad  {\cal N}(x')  \!\!\!\! \quad . 
\label{eq:intermediate-step}
\end{eqnarray} 
\\
where $\frac{1}{\Delta_{xx'}^2} $ is the Green's function of the biharmonic operator. The source ${\cal N}(x)$  is given by
\begin{equation}
{\cal N}(x) = \sum_{\alpha=1}^{N} q_{\alpha} \delta (x,x ^{\alpha}) - G(x) \!\!\!\! \quad . 
\label{eq:s1}
\end{equation}

Equation (\ref{eq:intermediate-step}) implies that the potential  $U(z)$, experienced by
an isolated disclination on a curved substrate with Gaussian curvature $G(z)$ satisfies the following biharmonic equation  
\begin{equation}\label{eq:GB5}
\nabla^4 U(z)= -G(z)  \!\!\!\! \quad ,
\end{equation}
and is therefore proportional to the Airy stress function $ \chi^{G}(z)$.

\begin{figure}[!t]
\centering
\includegraphics[width=\columnwidth]{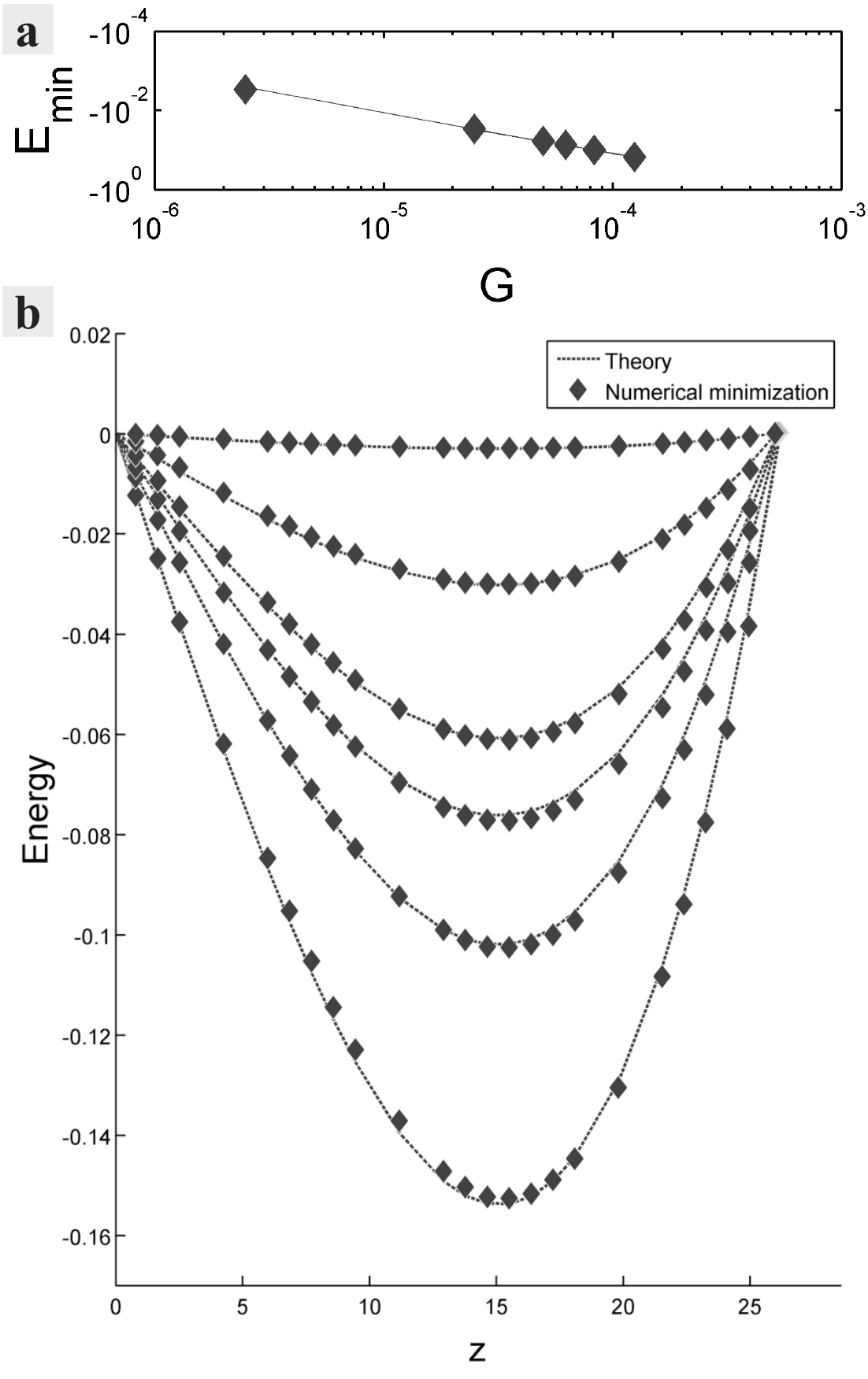}
\caption{Geometric potential for an isolated dislocation on surfaces of increasing Gaussian curvature, obtained in a similar way as the curves shown in Fig. 2. The scaling of the minimum energy as a function of the Gaussian curvature at the neck $G=\frac{1}{R_z R_{\phi}}$, is shown in {\bf (a)}, while a comparison of predicted geometric potentials (dashed lines) and energies found numerically  (solid rhombi)  is shown in {\bf (b)}}
\label{spherical_xtal}
\end{figure}

The geometric potential of a dislocation can be constructed by calculating the energy of a disclination dipole whose moment $qd_{i}=\epsilon_{ij} b_{j}$ is a lattice vector perpendicular to $\vec{b}$ that connects the two points of 5 and 7-fold symmetry {\it along the surface}. Upon taking the product of $d$ times the gradient in the z direction of the disclination energy $qU(z)$, including the factor $\sqrt{g_{zz}}= \cosh{(z/r_z)}$, we obtain the potential $\phi(z)$ and upon taking a second gradient the geometric force $f_z(z)$
\begin{equation}\label{eq:GB4}
f_z(z)=-\frac{1}{\sqrt{g_{zz}}}\partial_z \left[\frac{qd}{\sqrt{g_{zz}}}\partial_z U(z) \right] \!\!\!\! \quad 
\end{equation}
Once the identification $U(z)=\chi^{G}(z)$ and $qd=b$ is made, Eq. (\ref{eq:GB4}) becomes equivalent to Eq. (\ref{eq:a5}) . The energy $qU(z)$ of a disclination with topological charge $q$ is readily obtained from solving the Poisson equation $\nabla^2 U(z) = - \left( \rho (z) - \rho_H(z)\right)$, see Eq. (\ref{eq:metricw1}). Positive (negative) disclinations are repelled (attracted) from the neck of the catenoid (barrel) at $z\approx 0$ by the integrated background source $\rho (z) - \rho_H(z)$. 

We can also determine the geometric potential $\cal{I}$$(z)$ of an interstitial as the product of the local area expansion, $\delta A$, generated by the interstitial, times the local pressure $p(x)=-\nabla^2 \chi^G (x)$ generated by the curvature. The pressure (or equivalently the interstitial geometric potential)  is proportional to the right hand-side of Eq. (\ref{eq:metricw1}), eg. it is equal to the source of the Poisson equation including the crucial boundary term $\rho_{H}(x)$. To leading order in the surface deformation we find   
\begin{equation}\label{eq:I}
I(z) \sim \frac{Y b^2}{2 R_z R_{\phi}}\left( z^2-  \frac{H^2 }{12} \right)   \!\!\!\! \quad 
\end{equation}
where we assumed $\delta A \sim b^2 $.
 
\begin{figure}[!t]
\centering
\includegraphics[width=\columnwidth]{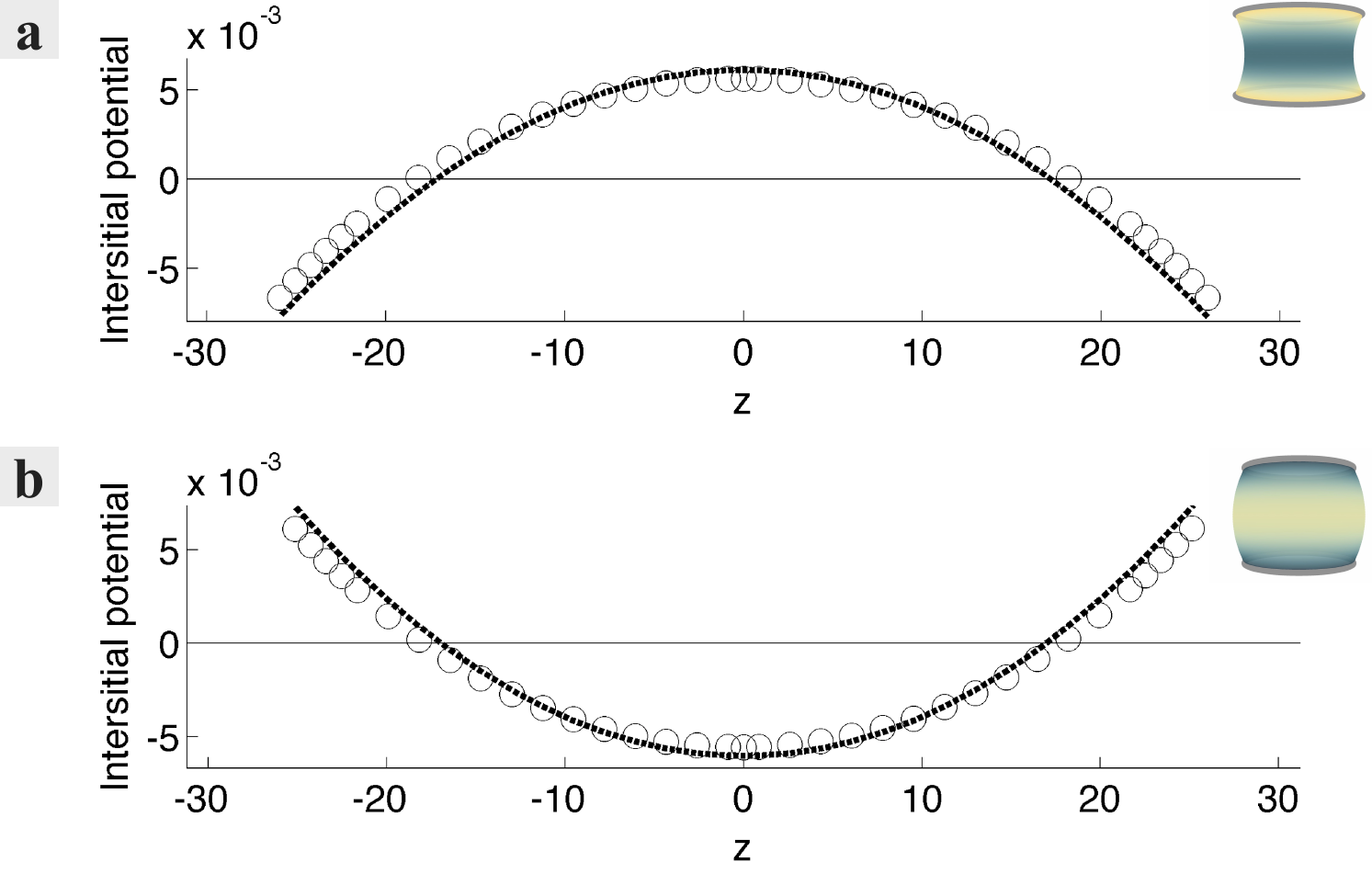}
\caption{Numerically calculated energy (black hollow circles) and predicted energy (dashed line)  of an interstitial particle on a capillary bridge having negative (a) and positive (b) curvature. The energy of an interstitial is given by $\rho(z) b^2$ and therefore provides a means to directly probe the geometric potential $\rho(z)$ on a surface.  }
\label{int1}
\end{figure}

The analytical expression for $\cal{I}$$(z)$ derived in Eq. (\ref{eq:I}) is plotted as a continuos line and compared to numerical results for catenoids  and barrels in Fig. \ref{int1}. Note that the only effect of switching the sign of the Gaussian curvature in going  from catenoids to barrels is the relative sign of $R_z$ and $R_{\phi}$. This determines the overall negative (positive) pre-factor of the parabolic potential for catenoids (barrels) in Eq. (\ref{eq:I}). 

The numerical data shown in Figure \ref{int1} are obtained from constrained energy minimization algorithms similar to the ones used to determine the dislocation potential. As shown before, we first determine numerically the interaction of the interstitial with its own image on a flat cylinder, and then subtract it from the total energy to isolate $\cal{I}$$(z)$ for a catenoid (panel a) and a barrel (panel b). 
The agreement is very good confirming the intuitive expectation that interstitials are repelled (attracted) from the neck of catenoids (barrels) where the frustrated lattice is compressed (stretched). Exactly the opposite trend is expected for vacancies which represent an area deficit. 

As we have illustrated, charge neutral dislocations can screen the Gaussian curvature of the capillary bridges as effectively as disclinations. We can estimate the threshold integrated Gaussian curvature required to make dislocations nucleation energetically favorable. On capillary bridges, this instability is a precursor to pleating: the formation of charge neutral grain boundary scars observed in the experiments of Ref. \cite{IVC:2010}. In what follows, we capture the onset of this instability by calculating the energetics of a pair of dislocations with opposite Burgers vectors oriented so that the 7s face the neck of the catenoid. 


Upon using Eq. (\ref{eq:a5}), the difference in energy, $\Delta E(z)$, between the curved crystal with the two dislocations at positions $\pm z$ and the curved crystal without any defects is
readily estimated  
\begin{equation}\label{eq:a1}
\Delta E(z)=\frac{Y b^2}{4 \pi} \ln{\left(\frac{2z}{b}\right)} + 2  b \int_{0}^z \sigma^G_{\phi \phi}(z') dz' \!\!\!\! \quad .
\end{equation}
The first term is the flat-space logarithmic interaction energy between two dislocations being pulled apart from the neck which approximates the interaction between dislocations to zero order. The second term accounts for the interaction of each of the two dislocation with the curvature. The energetic condition for the nucleation of a pair of oppositely oriented dislocations is
$\Delta E(z)<0$. 

Upon plugging $\sigma_{\phi\phi}^G$ into Eq. (\ref{eq:a1}) we obtain for $\Delta E \approx 0$:
\begin{equation}\label{eq:a3}
\left| \frac{ \pi}{9\sqrt{3}} \frac{H^2}{R_{\phi}R_{z}} \right|  \approx \frac{b}{H}  \ln{\left(\frac{H}{b}\right)} \!\!\!\! \quad ,
\end{equation}
The Burgers vector $b$ is typically of the order of the lattice spacing $a$ leading after simple rearrangements to 
\begin{equation}\label{eq:a4}
\left|\frac{2 \pi R_{\phi} H}{R_{\phi} R_{z}} \right| \approx \frac{a}{H}  \ln{\left(\frac{H}{a}\right)} \frac{18 \sqrt{3} R_{\phi}}{H}  \!\!\!\! \quad .
\end{equation}
The energetic condition can now be rewritten as a geometric condition on the threshold integrated Gaussian curvature:
\begin{equation}\label{eq:aa5}
\left| \int G \!\!\!\! \quad dA  \right| \approx  \frac{a}{H} \!\!\!\! \quad  \ln{\left(\frac{H}{a}\right)} \!\!\!\! \quad \frac{18 \sqrt{3} R_{\phi}}{H}.
\end{equation}
Note that for large samples the threshold integrated curvature is lowered: the instability is not suppressed but rather enhanced.

The geometric criterion for the nucleation of a dislocation pair, derived in Eq. (\ref{eq:aa5}), is qualitatively similar to heuristic estimates of the onset of pleating. Both instabilities are triggered by the energetic advantage derived from adding (or subtracting) rows of atoms where the curved crystal is most stretched (or compressed). Nucleating both dislocation pairs or pleats is associated with an energetic penalty that diverges slowly as the logarithm of the system size. By contrast, the energy $E_f$ of a stretched or compressed curved crystal characterized by a bond angle rotation $ \theta$ reads 
\begin{equation}\label{eq:a6}
E_f \sim Y  \theta ^2 H^2 \!\!\!\! \quad .
\end{equation}
Note that it diverges quadratically with the size of the system.

A single pleat is essentially a low angle grain boundary of {\it finite} length $H$ whose energy can be estimated as \cite{HL:1968}
\begin{equation}\label{eq:a6}
E_p \sim- Y b H \!\!\!\! \quad \theta \ln{\theta} \!\!\!\! \quad .
\end{equation}
As long as the curvature induced angular stretching (or compression) $\theta$ is small, it is energetically advantegeous to stretch (compress) bonds in the crystal rather than adding extra rows of atoms because  $E_f \sim \theta^2 < E_p$. 

Pleating becomes energetically favorable above a critical $\theta_c$
\begin{equation}\label{eq:a7}
\theta_c \sim -  \frac{b}{H} \!\!\!\! \quad \ln \!\!\!\! \quad \theta_c \!\!\!\! \quad .
\end{equation}
Equation (\ref{eq:a7}) can be solved self-consistently. Neglecting doubly logarithmic corrections leads to 
\begin{equation}\label{eq:a8}
\theta_c \sim  \frac{b}{H} \!\!\!\! \quad \ln \left(\frac{H}{b}\right) \!\!\!\! \quad .
\end{equation}
The curvature induced angular stretching (compression) is approximately equal to the integrated Gaussian curvature. This gives the criterion 
\begin{equation}\label{eq:a9}
\left| \int G dA \right| \sim \!\!\!\! \quad \frac{a}{H} \!\!\!\! \quad \ln \left(\frac{H}{a}\right)  \!\!\!\! \quad ,
\end{equation}
where the Burger vector is approximately one lattice spacing long. Note that Eq. (\ref{eq:a9}) 
provides an estimate of the same order as Eq. (\ref{eq:aa5}).

In this article, we have demonstrated that continuum elasticity can capture the energetics of defects on curved capillary bridges with free boundaries, as illustrated by the favorable comparison between our analytical
results and numerical energy minimizations of discrete spring models. Crucial to this agreement is the inclusion of a subtle geometric interaction between defects and a neutralizing background charge generated by the  boundary. Our results are in agreement with the experimental findings of Ref~\cite{IVC:2010}   and show that charge neutral pleats or dislocation pairs are effective in screening the Gaussian curvature on curved capillary bridges.

\section*{Acknowledgements}
We thank Paul Chaikin, Vinzenz Koning and Ariel Amir for useful discussions. 
V.V. Acknowledges funding from FOM and NWO.  
 This work was partially supported by the National Science Foundation (NSF) Materials
Research and Engineering Centers (MRSEC) Program at the University of
Chicago (DMR-0820054).


\begin{thebibliography}{30}

\bibitem{NelsonBook} D. R. Nelson, {\em Defects and Geometry in Condensed Matter Physics}, Cambridge University Press, Cambridge, 2002.

\bibitem{paulbook} P. Chaikin,  \& T. Lubensky,  {\it Principles of Condensed Matter Physics}, (Cambridge University Press, 1995). 

\bibitem{nelsonpeliti} D.R.~Nelson and L.~Peliti, Journal de Physique {\bf 48},  1085 (1987).    

\bibitem{BNT:2000} M. J. Bowick, D. R. Nelson and A. Travesset, {\em Phys. Rev. B} {\bf 62}, 8738 (2000). 

\bibitem{vv} V. Vitelli, J. B. Lucks \& D. R. Nelson,  Proc. Nat. Acad. Sci. USA {\bf 103}, 12323 (2006).

\bibitem{RMPV} A. M. Turner, V. Vitelli and D. R. Nelson, {\em Rev. Mod. Phys.} {\bf 82}, 1301 (2010).

\bibitem{RMPR} R. Kamien, {\em Rev. Mod. Phys.} {\bf 74}, 953 (2002). 

\bibitem{Santa} C. Santangelo, {V. Vitelli}, R. ~D. Kamien and D. ~R. Nelson, {\em Phys. Rev. Lett.} {99}, 017801 (2007).

\bibitem{BG:2009} M. J. Bowick and L. Giomi,  {\em Adv. Phys.}  {\bf 58}, 449 (2009).

\bibitem{mark} M. J. Bowick and Z. Yao, {\em Euro. Phys. Lett} {\bf 93},  36001 (2011).

\bibitem{mark2} L. Giomi and M. J. Bowick, {\em Eur. Phys. J. E}  {\bf 27}, 275 (2008).


\bibitem{Hex} A. Hexemer, V. Vitelli, E. J. Kramer and G. H. Fredrickson, {\em Phys. Rev. E} {\bf 76}, 051604 (2007).

\bibitem{Greg} G. M. Grason, {\em Phys. Rev. Lett.} {\bf 105}, 045502 (2010).

\bibitem{Tensor_book} J. H. Heinbockel, {\it Introduction to Tensor Calculus and Continuum Mechanics}, (Trafford, 2001). 

\bibitem{BB2003}  A. R. Bausch, M. J. Bowick, A. Cacciuto, A. D. Dinsmore, M. F. Hsu, D. R. Nelson, M. G. Nikolaides, A. Travesset and D. A. Weitz, {\em Science} {\bf 299}, 1716 (2003).

\bibitem{NatMat2005} P. Lipowsky, M. J. Bowick, J. H. Meinke, D. R. Nelson and A. B. Bausch, {\em Nat. Mater.} {\bf 4}, 407 (2005).

\bibitem{IVC:2010} W.T.M. Irvine,  V. Vitelli and P. Chaikin Nature {\bf 468}, 947 (2010).

\bibitem{HL:1968} J. P. Hirth and J. Lothe {\em Theory of Dislocations},  McGraw-Hill, New York, 1968.

\bibitem{PDM} A. P\`erez-Garrido, M. Dodgson \& M. Moore, Phys. Rev. B {\bf 56}, 3640 (1997).
































\end{thebibliography}
\end{document}